\documentclass[showpacs,prb,aps,floatfix,twocolumn]{revtex4}
\usepackage{epsfig}
\newcommand{\ql}{q_{{\bf l}}}
\newcommand{\be}{\begin{equation}}
\newcommand{\ee}{\end{equation}}
\begin{document}

\title{Planar pyrochlore: a strong-coupling analysis}
\author{Wolfram Brenig and Andreas Honecker}

\affiliation{Institut f\"ur Theoretische Physik,
Technische Universit\"at Braunschweig, 38106 Braunschweig,
Germany}

\date{November 21, 2001; revised February 19, 2002}

\begin{abstract}
\begin{center}
\parbox{14cm}{
Recent investigations of the two-dimensional spin-$1/2$ checkerboard lattice
favor a valence bond crystal with long range quadrumer order
[J.-B.~Fouet {\it et al.},
preprint cond-mat/0108070]. Starting from the limit of isolated quadrumers,
we perform a complementary analysis of the evolution of
the spectrum as a function of the inter quadrumer coupling $j$ using
both, exact diagonalization (ED) and series expansion (SE) by continuous
unitary transformation. We compute
(i) the ground state energy,
(ii) the elementary triplet excitations,
(iii) singlet excitations on finite systems and
find very good agreement between SE and ED.
In the thermodynamic limit we find a ground state energy substantially lower
than documented in the literature. The elementary triplet excitation is shown
to be gapped and almost dispersionless,
whereas the singlet sector contains strongly
dispersive modes. Evidence is presented for the low energy
singlet excitations in the spin gap in the vicinity of $j=1$ to result
from a large downward renormalization of local high-energy states.
}

\end{center}
\end{abstract}

\pacs{
75.10.Jm, 
75.50.Ee, 
75.40.-s  
}

\maketitle

\begin{figure}[t] 
\center{\psfig{file=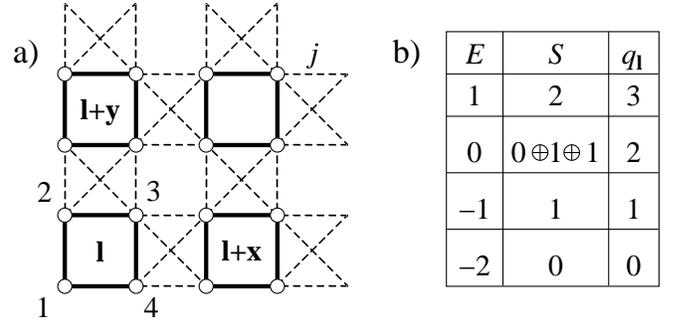,width=\columnwidth}}
\vspace{0.1cm}
\caption[l]{a) The checkerboard lattice. Spin-$1/2$ moments are located
on the open circles. Full (dashed) lines label the quadrumer bonds (bonds
corresponding to the expansion parameter $j$). b) Energy ($E$), spin ($S$),
and local $\ql$ quantum-number of a single quadrumer.}
\label{fig1}
\end{figure}

Quantum-magnetism in low dimensions has received considerable attention
recently due to the discovery
of numerous materials with spin-$1/2$ moments arranged in
chain, ladder, and depleted planar structures.
 Many of these systems exhibit
strongly gapped excitation spectra induced by various types of magnetic
dimerization and discrete symmetry breakings. Yet, low
energy collective spin dynamics may emerge from
materials with strong geometrical frustration leading to ground states with
near macroscopic degeneracy and possibly even zero temperature entropy
\cite{Ramirez96a,Lhuillier01a}. In this respect quantum spin
systems on the kagom\'e, and more recently on the
pyrochlore lattice are of particular interest.
While their classical counterparts have been studied in considerable
detail the role of quantum fluctuations in such systems
is an open issue. For the spin-$1/2$ kagom\'e lattice gapless singlet
excitations and a high density of singlet states in the singlet-triplet
gap have been established on finite systems by ED \cite{Lhuillier01a}.
Similar analysis of the pyrochlore quantum-magnet is severely constrained
by its three-dimensional structure.
Therefore, and as a first step, several
investigations\cite{Palmer01a,Elhajal01a,Moessner01a,Fouet2001} have
focussed
on the planar projection of the pyrochlore quantum-magnet, {\em i.e.}\ the
spin-$1/2$ checkerboard lattice of Fig.~\ref{fig1}a) for the case of
$j=1$. 

In particular, ED at $j=1$ has resulted in a sizeable spin gap and a large
number of in-gap singlet states\cite{Palmer01a,Fouet2001}. Moreover,
recent ED\cite{Fouet2001}
has given strong evidence in favor of a valence bond crystal (VBC) ground
state with long range order
in the $S=0$ quadrumers shown in Fig.~\ref{fig1}a). While this has 
been concluded from the quadrumer correlation at $j=1$ it emphasizes the need
for a perturbative investigation of the checkerboard magnet starting from
the limit $j\rightarrow 0$ in Fig.~\ref{fig1}a).
Low-order perturbation theory has been performed previously for the
checkerboard magnet\cite{Elhajal01a} however with the unperturbed
Hamiltonian chosen at variance with the VBC found in the ED\cite{Fouet2001}. 
Therefore, the purpose of this work is to explore the evolution of
the spin spectrum of the checkerboard spin-$1/2$ magnet
as a function of the inter quadrumer coupling $j$, contrasting
exact diagonalization (ED) against high-order series expansion (SE).

Written in a form adapted to the VBC symmetry breaking and normalized
to an overall unit of energy the Hamiltonian of
the checkerboard magnet reads
\begin{eqnarray}
\label{eq1}
H=\sum_{{\bf l}} 
&&\left[\frac{1}{2}({\bf P}^2_{1234{\bf l}}-{\bf P}^2_{13{\bf l}}
-{\bf P}^2_{24{\bf l}}) \right.
\nonumber\\
&& \left. +j\;({\bf P}_{34{\bf l}} {\bf P}_{12{\bf l}+{\bf x}}
+{\bf P}_{23{\bf l}}{\bf P}_{14{\bf l}+{\bf y}}) \right]
\\
\label{eq2}
=H_0  + && j\sum^{N}_{n=-N}T_n 
\end{eqnarray}
where ${\bf P}_{i\ldots j{\bf l}}={\bf S}_{i{\bf l}}+\ldots
+{\bf S}_{j{\bf l}}$ and ${\bf S}_{i{\bf l}}$ refers to spin-$1/2$
operators residing
on the vertices $i=1\ldots 4$ of the quadrumer at site ${\bf l}$, c.f.\
Fig.~\ref{fig1}a). $H_0$ is the sum over local quadrumer Hamiltonians
the spectrum of which, c.f. Fig.~\ref{fig1}b), consists of four equidistant
levels which can be labeled by spin $S$ and the
number of local energy quanta $\ql$.
$H_0$ displays an equidistant ladder spectrum
labeled by $Q=\sum_{{\bf l}} \ql$. The $Q=0$ sector is the {\em unperturbed}
ground state $|0\rangle$ of $H_0$, which is the VBC
of quadrumer-singlets. The $Q=1$-sector contains local $S=1$
single-particle excitations of the VBC with
$\ql=1$, where ${\bf l}$ runs over the lattice. At $Q=2$ the
spectrum of $H_0$ has a total $S=0,1$, or $2$ and is of multiparticle nature.
For $S=0$ at $Q=2$ it consists of one-particle singlets
with $\ql=2$ and two-particle singlets constructed from triplets
with $\ql=q_{{\bf m}}=1$ and ${\bf l}\neq {\bf m}$. Consequently the
perturbing terms $\propto j$ in (\ref{eq1}) can be written as a sum of
operators $T_n$ which {\em nonlocally} create(destroy) $n\geq 0$ ($n<0$)  
quanta within the ladder spectrum of $H_0$.

It has been shown recently\cite{Stein97,Mielke98,Knetter00} that problems
of type (\ref{eq2}) allow for high-order SE using a continuous unitary
transformation generated by the flow equation method of 
Wegner\cite{Wegner94}. The unitarily rotated effective
Hamiltonian $H_{\rm eff}$ reads\cite{Stein97,Knetter00}
\begin{eqnarray}
\label{eq3}
H_{\rm eff}=H_0+\sum^\infty_{n=1}\; j^n
\sum_{\stackrel{\mbox{\scriptsize $|{\bf m}|=n$}}{M({\bf m})=0}}
C({\bf m})\;T_{m_1}T_{m_2}\ldots T_{m_n}
\end{eqnarray}
where ${\bf m}=(m_1\ldots m_n)$ is an $n=|{\bf m}|$-tuple of integers,
each in a range of $m_i\in\{0,\pm 1,\ldots,\pm N\}$.
In contrast to $H$ of (\ref{eq1}), $H_{\rm eff}$ conserves the total number
of quanta $Q$. This
is evident from the constraint $M({\bf m})=\sum^n_{i=1} m_i=0$.
The amplitudes $C({\bf m})$ are rational numbers computed from the
flow equation method \cite{Stein97,Knetter00}. 

Previous applications of this method to spin models have
been confined to dimer systems\cite{Knetter00}
where $N\leq 2$. Here we report on its first application using a quadrumer
basis where $N\leq 6$. Explicit tabulation\cite{HttpRef} of the $T_n$
shows that for the checkerboard magnet $N=4$.
Because the quadrumer
basis consists of 16(4) states(spins) rather than only 4(2) states(spins)
as for dimer systems, the maximum order achievable by a quadrumer
expansion is lower in general than for a dimer expansion. Here
we present results up to $O(7)$ for the ground state energy
and the elementary triplet
as well as up to $O(6)$ for the $Q=2$ singlets. The $C({\bf m})$-table is
available on request\cite{HttpRef}.

$Q$-conservation leads to a ground state energy of
\begin{eqnarray}
\label{eq4}
E_g=\langle 0|H_{\rm eff}|0\rangle \, .
\end{eqnarray}
Evaluating this matrix element on clusters with periodic boundary
conditions, sufficiently large not to allow for wrap around at
graph-length $n$ one can obtain SEs valid to $O(n)$ in the
thermodynamic limit, {\em i.e.}\ for systems of infinite size.

$Q$-number conservation guarantees the $Q=1$-triplets
to remain genuine one-particle states. {\em A priori } single-particle 
states from sectors with $Q > 1$
will not only disperse via $H_{\rm eff}$, but can decay into
multi-particle states. {\em A posteriori } however such decay
may happen only at high order in $j$ leaving the excitations almost true
one-particle states. The dispersion of the single-particle excitations is
\begin{eqnarray}
\label{eq5}
E_\mu({\bf k})=\sum_{lm} t_{\mu,lm} e^{i (k_x l+k_y m)}
\end{eqnarray}
where
$t_{\mu,lm}=\langle \mu,lm|H_{\rm eff}|\mu,00\rangle
- \delta_{lm,00}E^{obc}_g$
are hopping matrix elements from site $(0,0)$ to site
$(l,m)$ for a quadrumer excitation $\mu$
inserted into the unperturbed ground state.
For the thermodynamic limit $t_{\mu,lm}$ has to be evaluated on clusters with
open boundary conditions large enough to embed all linked
paths of length $n$ connecting sites $(0,0)$ to $(l,m)$ at $O(n)$ of
the perturbation. $E^{obc}_g=\langle 0|H_{\rm eff}|0\rangle$ on the
$t_{\mu,00}$-cluster. 

To assess the convergence of the SEs, as well as to identify the remaining
excitation spectrum, we will contrast the SEs with ED of finite systems.
For this purpose eqns.~(\ref{eq4},\ref{eq5}) apply equally well, with however
the cluster topologies set by the finite systems. We emphasize, that
for some of the SEs generated by (\ref{eq3}) on finite systems
we were also able to compute standard SEs using a completely
different code~\cite{Pert}.
In all cases we found both SEs to agree which serves as an
independent check.

\begin{figure}[t]
\center{\psfig{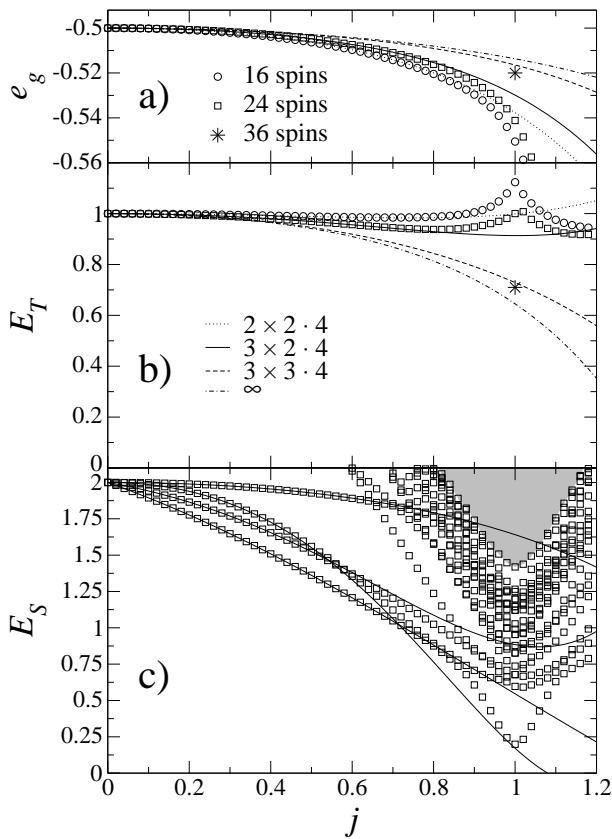}}
\vspace{0.1cm}
\caption[l]{a) Ground state energy per spin, b) spin gap and 
c) singlet excitations with ${\bf k} = {\bf 0}$ as a function of $j$.
Lines correspond to the series, symbols to exact diagonalization.
The numerical data at $j=1$ for 36 spins is taken from
Ref.~\protect{\onlinecite{Fouet2001}}. The grey shaded area in panel c)
indicates that our numerical data is not complete in this region.}
\label{compare}
\end{figure}

Fig.~\ref{compare} summarizes our results for the ground state
energy {\em per spin} $e_g(j)$, the spin gap $E_T(j)$, and the
spectrum of the low lying singlets. The figure compares the SEs
with our ED data on $2(3)\times 2\cdot 4$ systems, {\it i.e.}\ 16(24)
spins, as a function of $j$,
as well as available ED data\cite{Fouet2001} on $3\times 3\cdot 4$
sites, {\it i.e.}\ 36 spins, at $j=1$. SEs are shown both
for the {\em identical} system sizes and also for the thermodynamic limit
to $O(j^7)$ for $e_g$ and $E_T$ as well as to $O(j^6)$
for the $Q=2$-singlets.  All SE data refers to the
actual series, no Pad\'e continuations have been applied.

First we focus on the ground state energy in Fig.~\ref{compare}a).
In the thermodynamic limit we find
\begin{eqnarray}
\label{eq6}
e^\infty_g(j)=&&-\frac{1}{2}  - \frac{j^2}{96} - \frac{j^3}{768} - 
  \frac{451\,j^4}{387072} - \frac{127\,j^5}{338688}
\nonumber\\  
&& - \frac{865153\,j^6}{3072577536} - 
  \frac{24293225710331\,j^7}{178396309802188800}
\end{eqnarray}
which is lower than the decoupled quadrumer-value of $e_g(0)=- 1/2$
for all $j\leq 1$ and yields $e^\infty_g(1)\approx -0.513677$. This
is also lower than the value of $\tilde{e}^\infty_g(1)\approx -0.49622$
which is {\em above} the bare quadrumer limit and has been reported from
a different third-order SE
starting from decoupled {\em tetrahedra}\cite{Elhajal01a}. While
the SEs are not related to a variational principle the preceding
is consistent with the decoupled quadrumers
to incorporate the proper ground state
correlations for $j\leq 1$ rather than the tetrahedral limit. This is
confirmed on the finite systems in Fig.~\ref{compare}a), where practically
quantitative agreement is found for $j\lesssim 0.7$ between ED and SE
for $2(3)\times 2\cdot 4$ sites. At $j\sim 1$ very small deviations occur,
which however decrease with system size. They are $\sim 2$\% on the
16(24)-spin lattices and only 0.7\% on the $3\times 3\cdot 4$ site 
system\cite{Fouet2001}.
Regarding the SEs this agreement is remarkable also,
since the high-order decrease of SE-coefficients\cite{HttpRef} tends to
be slowed down on finite geometries. On the $2(3)\times 2\cdot 4$ lattice
we observe a kink in the $e_g(j)$ as obtained from the ED
at $j\approx 1$ which suggests a change of the nature of the ground state
on these
systems in the vicinity of $j=1$. This is probably a special property
of the 16(24)-spin lattices\cite{Fouet2001}.

Next we discuss the elementary triplet excitations. The SE to $O(j^7)$
in the thermodynamic limit for
the triplet dispersion reads
\begin{eqnarray}
\label{eq7}
&&E^\infty_T({\bf k},j) = 1 - \frac{7\,j^2}{36} - \frac{41\,j^3}{864} 
- \frac{329887\,j^4}{6531840} - \frac{580309487\,j^5}{21946982400}
\nonumber\\
&& -\frac{16957803829\,j^6}{790091366400} - 
\frac{7822020675129119\,j^7}{557488468131840000} 
- \left( \frac{53\,j^5}{1036800}  \right.
\nonumber\\
&&\left.  + \frac{59527\,j^6}{1306368000}
+ \frac{74504581093\,j^7}{948109639680000} \right) \,
\left( \cos (k_x) + \cos (k_y) \right) 
\nonumber\\
&& + \left( \frac{1679\,j^6}{32659200}
+ \frac{2039741\,j^7}{438939648000} \right)\,\cos (k_x)\,\cos (k_y) \, .
\end{eqnarray}
Remarkably, the hopping amplitude is exceedingly small and sets in only
at $O(j^5)$ hinting at a fairly
extended polarization cloud necessary in order to allow for triplet
motion\cite{Knetter00b}. From (\ref{eq7}) the spin gap occurs at
${\bf k}={\bf 0}$.

The spin gap is shown in Fig.~\ref{compare}b)\cite{complement}.
Again, on finite systems
of $2(3)\times 2\cdot 4$ sites and for $j\lesssim 0.7$ the agreement
between ED and $O(j^7)$-SE is very good. At $j=1$ and for the
SEs shown the absolute value of the relative difference to the
numerically exact values for the spin gap improves from 11\%, to 8.6\%,
to 2.7\% on passing from 16, to 24, to 36 spins. A clear cusp is observable
in $E_T(j)$ close to $j=1$ on the $2(3)\times 2\cdot 4$ systems.
This is closely related to the previously anticipated change
in the nature of the ground state on the 16(24)-spin systems
at $j\approx 1$ and is not captured by the SE. 
While no $j$-scan of $E_T$ is available on $3\times 3\cdot 4$ sites
it is remarkable that at $j=1$ the SE agrees better with ED
for 36 spins than for 16(24) spins and that the SE behaves qualitatively
different, both on the $3\times 3\cdot 4$ lattice {\em and} in the
thermodynamic limit as compared to the $2(3)\times 2\cdot 4$ lattices.
{\it I.e.}, on the larger systems the gap decreases monotonously, while it
increases beyond $j\gtrsim 1$ on the smaller two. This suggests an
absence -- or strong reduction -- of the discontinuities close to $j=1$ on
larger systems improving the convergence of the SE.

Finally we discuss the low energy singlet excitations. At $j\ll 1$
they result from the $Q=2$ sector which involves solving a two-particle
problem coupled to a one-particle problem. We have found
the mixing between the one- and two-particle states in the $S=0$, $Q=2$
sector to be very weak. In fact, on the $2\times 2\cdot 4$ system the mixing
{\em vanishes} identically in the ED as well as the SE up to $O(j^7)$.
For all other systems analyzed by SE including the thermodynamic limit,
the mixing sets in only at $O(j^5)$. Therefore the $Q=2$
one-particle singlet remains an almost true one-particle state
even at finite $j$. In addition its dispersion $E^\infty_{S,1pt}$ 
up to $O(j^4)$ is determined {\em exactly} by (\ref{eq5}). In the
thermodynamic limit we get
\begin{eqnarray}
\label{eq8}
E^\infty_{S,1pt}&&({\bf k},j) = 2 - \frac{5\,j^2}{12} - \frac{11\,j^3}{96} 
- \frac{227\,j^4}{3024} 
- \left( \frac{j^2}{4} + \frac{j^3}{16} \right.
\nonumber\\
&& \left. + \frac{23\,j^4}{2304} \right)
\left( \cos (k_x) + \cos (k_y) \right)  - 
  \frac{3\,j^4}{512} \,( \cos (2\,k_x) 
\nonumber\\
&&  + \cos (2\,k_y)) 
+ \frac{53\,j^4}{1152} \,\cos (k_x)\,\cos (k_y) \, .
\end{eqnarray} 
In sharp contrast to (\ref{eq7}) this low energy singlet is strongly
dispersive. Its gap resides at ${\bf k}={\bf 0}$.

Fig.~\ref{compare}c) shows the evolution of all eigenvalues of $H$
from (\ref{eq1}) in the singlet sector
on the $3\times 2\cdot 4$ system at ${\bf k}={\bf 0}$ as well as the
spectrum of $H_{\rm eff}$ at $O(j^6)$ in the $S=0$, $Q=2$ sector. For
$j\lesssim 0.7$ the agreement between ED and SE is satisfying.
Similar agreement is found at the remaining ${\bf k}$-vectors.
The lowest $S=0$ excitation can be identified with a singly degenerate
$Q=2$ two-triplet excitation. The binding energy for this
two-particle state is approximately proportional to $j$ with a
coefficient of order unity. Starting at $j\gtrsim 0.7$ a large number of
singlets from higher $Q$-sector enter the low energy spectrum.
At $j=1$ several of these singlets reside in the spin gap. 
This is consistent with ED at $j=1$ in Refs.~\onlinecite{Palmer01a,Fouet2001}.
The singlet gap occurs at ${\bf k}={\bf 0}$. Preliminary analysis indicates
that the lowest singlets have a substantial overlap with the $Q=4$ sector.
The numerical spectrum in Fig.~\ref{compare}c)
is very suggestive of a softening in the singlet sector in the
vicinity of $j=1$. If this is a feature particular only to the finite
systems is an issue in clear need of further study.
Close to $j=0.7$ a near level crossing occurs, with the lowest state
of the SE for $j\gtrsim 0.7$ corresponding to a continuation of the
$Q=2$ one-particle singlet. 

\begin{figure}[t]
\center{\psfig{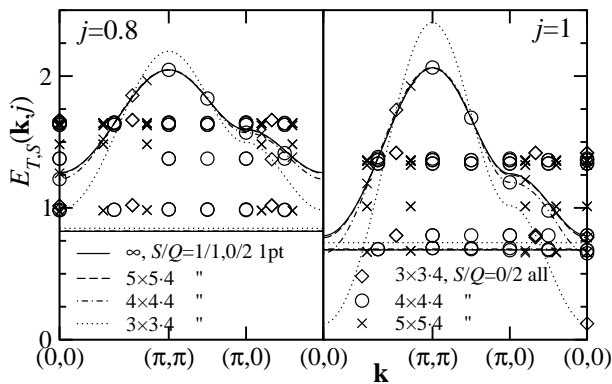}}
\vspace{0.1cm}
\caption[l]{System size dependence of the dispersion from SE for
$Q=1$, $S=1$ and $Q=2$, $S=0$. Lines: $E_T({\bf k},j)$ (almost
nondispersive) and $E_{S,1pt}({\bf k},j)$ (dispersive).
Symbols: all eigenvalues of the $Q=2$, $S=0$ sector.}
\label{finsiz}
\end{figure}

Finally we speculate on the finite size dependence to be expected for
systems presently inaccessible to ED focusing on the $Q=1$ triplet-
and the $Q=2$, $S=0$ singlet-sector by using SE instead.
In Fig.~\ref{finsiz} we show the spectrum of
$H_{\rm eff}$ for these two sectors at $O(j^5)$ for $j=0.8$ and $1.0$
along a path in the Brillouin zone (BZ) on four systems ranging from
$N=3\times 3\cdot 4$ sites up to the thermodynamic
limit. As we have pointed out SEs will converge more rapidly on larger
systems. Therefore, at $j=0.8$ the SE results of Fig.~\ref{finsiz}
will be rather well converged while at $j=1$ they
will be qualitatively correct. Fig.~\ref{finsiz} includes
the one-particle triplet and singlet dispersions,
$E_T({\bf k},j)$ and $E_{S,1pt}({\bf k},j)$ with ${\bf k}$
varying continuously along the path in the BZ. For the latter curve
$E_{S,1pt}({\bf k},j)$ has been evaluated at $O(j^5)$. Moreover
in the thermodynamic limit and for $Q=2$ only $E_{S,1pt}({\bf k},j)$
is depicted.
First Fig.~\ref{finsiz} illustrates the difference in
strength of the dispersion between the elementary triplet and
almost one-particle singlet from the $Q=2$-sector. Second the
figure shows that both, the elementary triplets and the
corresponding, almost two-particle singlets with $Q=2$ display
only a small downward shift with increasing system size. However,
the spectrum of the almost one-particle singlet with $Q=2$
experiences a strong reduction of its band-width including an upward
shift of its gap. In particular, at $j=1$ the SE suggests that
the $Q=2$-singlets are above the spin gap for sufficiently
large systems and that only singlets from sectors with $Q>2$
might reside in the spin gap\cite{llsRemark}.
This underlines the need, both,
for larger systems in ED and larger $Q$-values in SE studies of the
singlet sector.

In conclusion, by comparing ED and SE we find the decoupled
quadrumer-limit to be an excellent starting point for a perturbative
treatment of the spin-$1/2$ checkerboard magnet even at $j \approx 1$.
The remarkably fast convergence of the series for the ground state and
the elementary triplet excitation up to at least $j=1$ indicates
that the points $j=0$ and $1$ belong to the same phase. This is consistent
with the strong VBC correlations observed\cite{Fouet2001} at $j=1$. 
We have further shown the
elementary triplets of the checkerboard magnet to be almost nondispersive.
Up to intermediate $j$ the elementary singlet
excitations are almost nondispersive two-particle states from the $Q=2$
sector residing above the spin gap. On finite systems and in the vicinity
of $j = 1$, we find a strong downward shift of singlets from higher $Q$
sectors, entering the spin gap. In future work it would be interesting
to study these higher $Q$-sectors by SE and to obtain $j$-dependent ED
spectra on larger systems, extending in particular the 36-site results of
Ref.~\onlinecite{Fouet2001} to $j\ne 1$.

{\bf Acknowledgments: } We are grateful to C.\ Lhuillier, P.\ Sindzingre,
and J.-B.\ Fouet for helpful discussions and for communicating to us
the results of Ref.~\onlinecite{Fouet2001} prior to publication. W.B.\
also acknowledges the kind hospitality of the LPTL at Universit{\'e}
Paris VI where part of this work has been initiated. This
research was supported in part by the DFG under Grant BR 1084/1-2.


\begin{thebibliography}{10}
\bibitem{HttpRef}
These results are
available on request from the authors.
\bibitem{llsRemark}
Since the VBC picture implies a two-fold ground state
degeneracy\cite{Fouet2001} at $j=1$, we expect
at least one low-lying singlet excitation to come from these
higher $Q$-sectors.
Due to $Q$-conservation it is possible that the mixing of
this state with the ground state as well as the convergence
problems usually associated with this can be avoided in the present case.
\end{thebibliography}
\end{document}